\documentclass[doublecol]{epl2} 

\usepackage{graphicx,bbm,setspace,amssymb,amsfonts,amsmath,mathtools,mathrsfs,stmaryrd,amsthm,lingmacros,enumitem,algorithm,nicefrac,bm,caption,subcaption,etoolbox,comment,url,booktabs,xcolor}

\usepackage{dcolumn}

\newtheorem{thm-defn}[theorem]{Theorem/Definition}

\theoremstyle{definition}

\theoremstyle{remark}

\title{Portfolio diversification with varying investor abilities}
\shorttitle{Portfolio diversification with varying investor abilities} 

\author{N. James\inst{1} \and M. Menzies \inst{2}}
\shortauthor{N. James \and M. Menzies}

\institute{                    
  \inst{1} School of Mathematics and Statistics, University of Melbourne, Victoria 3010, Australia\\
  \inst{2} Yanqi Lake Beijing Institute of Mathematical Sciences and Applications, Beijing 101408, China
}

\abstract{
We introduce new mathematical methods to study the optimal portfolio size of investment portfolios over time, considering investors with varying skill levels. First, we explore the benefit of portfolio diversification on an annual basis for poor, average and strong investors defined by the 10th, 50th and 90th percentiles of risk-adjusted returns, respectively. Second, we conduct a thorough regression experiment examining quantiles of risk-adjusted returns as a function of portfolio size across investor ability, testing for trends and curvature within these functions. Finally, we study the optimal portfolio size for poor, average and strong investors in a continuously temporal manner using more than 20 years of data. We show that strong investors should hold concentrated portfolios, poor investors should hold diversified portfolios; average investors have a less obvious distribution with the optimal number varying materially over time. 
}

\begin{document}

\maketitle


\section{Introduction}

The present day presents an unprecedented challenge for portfolio managers, both retail and institutional. Even if we are not strictly speaking in a financial crisis, the contemporary market period is rather unprecedented. Following two recent market crises caused by COVID-19 \cite{Cheng2022_volatility,Priscilla2022} and the Russian invasion of Ukraine \cite{Fang2022,Kele2023} and a strong bull market in between, we have again begun to observe patches of an economic slowdown \cite{deloitte_slowdown}. The International Monetary Fund (IMF) and several policymakers have warned of a sharp economic recession, termed a ``hard landing'' \cite{cnbc_hardlanding}. Inflation has been high since 2020 \cite{pew_inflation}, and despite continual increases in interest rates \cite{ap_interestrates}, central banks believe it remains too high and may continue to take action \cite{FT_inflation}.

Beyond heightened inflation and interest rates, several aspects of this current period are quite unusual. First, there has been a clear disconnect between fixed income and equity markets, unlike prior periods of financial distress \cite{blackrock_disconnect}. Second, the ongoing Ukraine war has greatly impacted energy prices \cite{rand_ukraine}, while the usually reliable technology sector has instituted multiple rounds of retrenchments \cite{tech_layoffs}. On the other hand, generative artificial intelligence is buoying numerous equities in the same sector and creating excitement throughout the broader economy for its potential opportunities \cite{McKinsey_AI}. Such changes have caused great contradiction within equity markets. New asset classes such as cryptocurrencies have also seen great volatility and contradiction, with  highs in 2021 dispersed around significant exchange crashes \cite{Forbes_Bitmex,NYT_FTX}. The equity market volatility exhibited both clearly defined market crises as well as the unusual present day period motivate our study of optimal portfolio composition during diverse conditions.

Since Markowitz' mean-variance model \cite{Markowitz1952,Sharpe1966}, there has been a substantial body of work on portfolio optimisation. Different approaches include statistical mechanics \cite{Zhao2016,Li2021_portfolio,james2021_portfolio}, clustering \cite{Iorio2018,Len2017,james_arjun}, fuzzy sets \cite{Tanaka2000,Ammar2003}, networks \cite{james_georg}, regularisation \cite{Fastrich2014,Li2015,Pun2019}, Bayesian approaches \cite{james2021_spectral,james2021_MJW} and multiobjective optimisation \cite{Lam2021}. For further details, a review of such techniques for portfolio optimisation was conducted by \cite{Milhomem2020}. In particular, a significant challenge in portfolio selection is due to the fact that many standard portfolio conditions yield non-convex and NP-hard optimisation problems \cite{Shaw2008}. Perhaps the simplest and most common example would be a cardinality constraint \cite{Anagnostopoulos2011}, common among retail and institutional investors alike \cite{Russellpolicy}, where a portfolio should only contain a limited number of assets to minimise complexity and trading costs. \cite{Jin2016} provided a review of typical constraints in an asset allocation problem, as well as advances in algorithmic procedures to tackle non-convex problems \cite{Meghwani2017,Lwin2014}.

This paper departs from the traditional view of portfolio optimisation with three inspirations. First, we acknowledge portfolio optimisation as a difficult procedure, especially over non-convex sets. Second, we consider a growing body of work that it is frequently difficult to beat equally weighted portfolios \cite{DeMiguel2007,Farago2022}. Third, we address the idea of ``investor skill,'' where we study the optimal cardinality of portfolios for investors with varying levels of success in picking ``winning'' securities. Indeed, we cannot guarantee that all or many portfolio managers would be equally skilled or knowledgeable to have the most effective algorithmic methods at their selection (whose performance over simpler equal weightings may be doubtful). Thus, in this paper, we specifically investigate two aspects of portfolio selection: what the distribution of portfolio outcomes looks like across different cardinalities $k$ and different levels of investor skill. We depart from traditional optimisation methodologies and instead analyse portfolio performance by means of a novel simulation-based framework inspired by techniques from statistical mechanics, in which we record the top 10\%, bottom 10\% and median performance of various randomly chosen equally weighted portfolios. In the two subsequent sections, we study this on a year-by-year and a continual time-varying basis, respectively.

We must acknowledge the influence of statistical physics, econophysics, and time series analysis on this work. In financial markets, these methodologies have been applied to a wide range of asset classes such as equities \cite{james2022_stagflation,Wilcox2007,Alves2020,James2023_financialcrises}, foreign exchange \cite{Ausloos2000}, cryptocurrencies \cite{Gbarowski2019,james2021_crypto2,DrodKwapie2022_crypto,DrodWtorek2022_crypto,Drod2020,James2023_cryptoGeorg,Drod2023_crypto2,DrodWtorek2023_crypto} and debt-related instruments \cite{Driessen2003}. Such methods from applied mathematics have been used in a variety of other disciplines including epidemiology \cite{jamescovideu,Manchein2020,Li2021_Matjaz,Blasius2020,james2021_TVO,Perc2020,Machado2020,james2021_CovidIndia,james2023_covidinfectivity,Sunahara2023_Matjaz}, environmental sciences \cite{james2022_CO2,Khan2020,Derwent1995,james2021_hydrogen,Westmoreland2007,james2020_Lp,Grange2018,james2023_hydrogen2,Libiseller2005}, crime \cite{james2022_guns,Perc2013,james2023_terrorist}, and other fields \cite{Clauset2015,james2021_olympics}.

In the subsequent sections, we first introduce the data analysed in this study. Then, we conduct our annual analysis, where distributions of portfolios based on size and skill level are investigated on a year-by-year basis. The next section implements our continuous analysis, where distributions are studied on a rolling 90-day basis. Finally, we conclude and summarise the main implications of this work.


\section{Data}

Our data consists of $N=370$ US equities with daily trading data from 04/08/1999 to 04/08/2022. We have chosen all S\&P 500 equities currently in existence that have at least 20 years of history, and study their behaviour during a range of conditions. Biases do exist in our dataset. Survivorship bias and daily data periodicity may impact expected equity returns and volatility, respectively, while only sampling US equities may influence our findings regarding portfolio diversification.

We partition our data into one-year equal length periods. Throughout our analysis, we will compute the Sharpe ratio of a portfolio, defined by
\begin{align}
\label{eq:Sharpeobjectionfn}
 \frac{\sum^{k}_{i=1} w_{i} R_{i} - R_f}{ \sqrt{\boldsymbol{w}^{T} \Sigma \boldsymbol{w}}  }.
\end{align}
In (\ref{eq:Sharpeobjectionfn}) above, $R_i$ is the return of a stock over some candidate period, $\Sigma$ is the covariance matrix between stocks over the period, $w_i$ is the weight of the stock in a portfolio, and $R_f$ is the risk-free rate. In this paper, we only use equally weighted portfolios, with all $w_i$ set to $\frac{1}{k}$ for a size $k$ portfolio, and a risk-free rate of 0. The Sharpe ratio is a measure of the risk-adjusted returns of a portfolio that seeks to reward portfolios with high returns, while simultaneously penalising excessive variance.

\section{Annual analysis}
\label{sec:annualanalysis}

In this first section, we analyse the distributions of Sharpe ratios for various cardinalities on a year-by-year basis. That is, we set a period $P=252$, the number of trading days in a year, and consider non-overlapping yearly periods of $P$ trading days that partition our data into 23 years.

The following analysis is performed for every year-length period: let $k$ range from 10 to 100. For each $k$, we sample 1000 different equally weighted portfolios of $k$ stocks from our collection of $N$. Effectively, this samples from the full space $\mathcal{C}_{N,k}$ of all $N \choose k$ equally weighted portfolios. We record the empirical median as well as top 10\% and bottom 10\% quantile values of the Sharpe ratio. That is, we empirically analyse the random variable
\begin{align}
\label{eq:SharpeRV}
\mathcal{S}_k=\text{Sharpe}: \mathcal{C}_{N,k} \to \mathbb{R}
\end{align}
and interpret its quantiles as a function of $k$. Let $Q_{\mathcal{S},k}$ be the associated quantile function of the random variable $\mathcal{S}_k$. We record and display quantile values $Q_{\mathcal{S},k}(q)$ for $q \in \{ 0.1, 0.5, 0.9\}$ in Figure \ref{fig:Portfolio_k_curves}.

We regard $q=0.1,0.5,0.9$ as a representative Sharpe ratio obtained by a low-skill, medium-skill and high-skill investor respectively. For each of these quantile categories, we ask: which value of $k$ offers the best performance? We answer this with two approaches.

First, we compute the \emph{raw optimum} $k_0$ for each skill category (corresponding to quantiles $q$) and yearly period under consideration. This is simply the value of $k$ that gives the highest Sharpe ratio for a given period and quantile. Formally, it is the argmax of the following function:
\begin{align}
\label{eq:nonpenalisedmax}
   \{10,11,...,100\} \to \mathbb{R}; k \mapsto Q_{\mathcal{S},k}(q).
\end{align}

Second, we compute a \emph{penalised optimum} $\hat{k}$. This is pertinent and appropriate primarily for the lowest quantile $q=0.1$. Our motivation is as follows: we observe that for this lowest quantile, that representative Sharpe ratio values tend to increase near-uniformly with increasing $k$. However, it is not satisfactory simply to always suggest selecting the largest $k$, as increases in portfolio size bring added costs and complexity. Thus, we impose a slight data-driven penalty for this increase in size.

Specifically, we fit a univariate linear regression whose predictor ($x$) values are the range of $k$ from 10 to 100 and whose response ($y$) values are the representative Sharpe ratios $Q_\mathcal{S}(q)$ as a function of $k$. Simply put, our motivation here is to see how the performance of a typical portfolio at a specified level of skill (low-skilled, median or high-skilled) changes with portfolio size. Let $\hat{\beta}^1_1$ be the determined slope of the regression line. Then $\hat{\beta}^1_1$ gives the average increase of the representative Sharpe with $k$ across $10 \leq k \leq 100$. If the representative Sharpe is increasing ahead of this line, we consider that an efficient and worthy increase in Sharpe ratio with $k$ - ``better than average.'' Thus, we impose a penalty of $\hat{\beta}^1_1$ for each increment in $k$, and seek to maximise the penalised quantile function
\begin{align}
\label{eq:penalisedmax}
  \{10,11,...,100\} \to \mathbb{R};  k \mapsto Q_{\mathcal{S},k}(q) - k \hat{\beta}^1_1.
\end{align}
Maximising this quantity has an interpretation in terms of the mean value theorem. When (\ref{eq:penalisedmax}) is maximal, one should understand the derivative to be zero. Indeed, if $Q_\mathcal{S}$ were a continuous function, then $\hat{k}$ would occur at the point where $Q_\mathcal{S}' =0$,  where the instantaneous growth rate of a concave down function follows below the average growth rate across the entire interval, guaranteed by the mean value theorem. Of course, we adopt this maximum in our discrete setting, and we use a linear fit estimator of the line slope $\hat{\beta}^1_1$ rather than simply the line slope connecting the endpoints for more robustness against outliers.

In Tables 1,2 and 3 of the supplementary material, we record the optimum $k_0$ for each quantile $q=0.1,0.5,0.9$ and year under consideration, and the penalised optimum $\hat{k}$ for $q=0.1.$ In addition, we make sure to compare the raw optimum and penalised values in Figure 2 of the supplementary. There, we display a scatter plot of the deviation in representative Sharpe ratios $\Delta=Q_{\mathcal{S},k_0}(q) - Q_{\mathcal{S},\hat{k}}(q)$ against the raw optimal Sharpe ratio $Q_{\mathcal{S},k_0}(q)$, specifically for $q=0.1$, over all our yearly intervals. This deviation $\Delta$ is by definition always positive, and we can see that for most years, it is relatively small, indicating that in performance terms, little is lost by using the penalised optimum in the case of a bottom 10\% portfolio.

Finally, motivated by Figure \ref{fig:Portfolio_k_curves}, we notice that the functions $k \mapsto Q_{\mathcal{S},k}(q)$ exhibit one of two behaviours: apparent linearity or some degree of curvature. Thus, we perform a regression analysis, where for each quantile $q$ and time interval (year) under consideration, we fit both a linear and quadratic model. For example, a quadratic fit takes the following form:
\begin{align}
\label{eq:quadraticmodel}
Q_{\mathcal{S},k}(q) \sim \beta_0 + \beta_1 k + \beta_2 k^2,
\end{align}
with $\beta_i$ selected by least squares regression, and analogously for a linear model. In Tables 1,2,3 of the supplementary material, we record statistical parameters for each of these models, and determine the selected model. This offers a data-driven answer to the question of whether or not significant curvature is observed when comparing the change of $Q_{\mathcal{S},k}(q)$ relative to $k$. We also plot linear and quadratic fits for selected years in Figure \ref{fig:Portfolio_regressions}.

\subsection{Annual analysis results}

\begin{figure*}
    \centering
    \begin{subfigure}[b]{0.32\textwidth}
        \includegraphics[width=\textwidth]{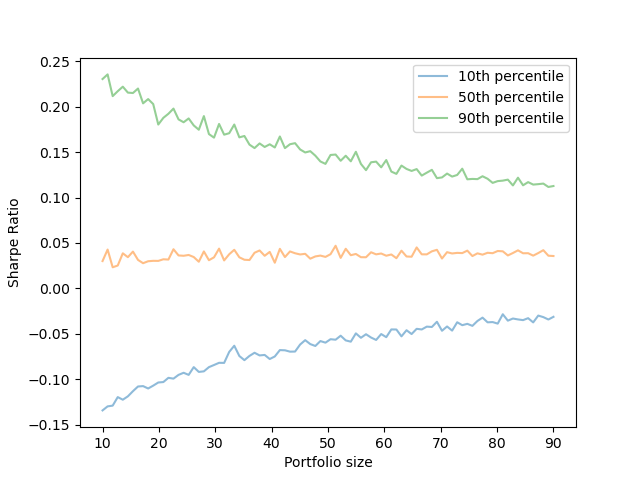}
        \caption{} 
        \label{fig:Portfolio_bear_1}
    \end{subfigure}
    \begin{subfigure}[b]{0.32\textwidth}
        \includegraphics[width=\textwidth]{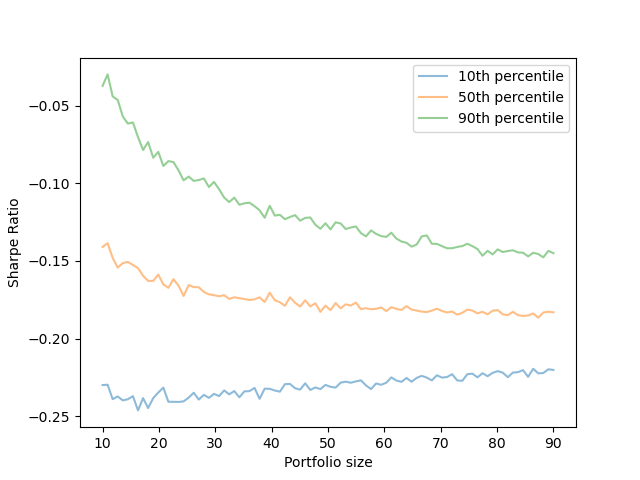}
        \caption{} 
        \label{fig:Portfolio_bear_2}
    \end{subfigure}
        \begin{subfigure}[b]{0.32\textwidth}
        \includegraphics[width=\textwidth]{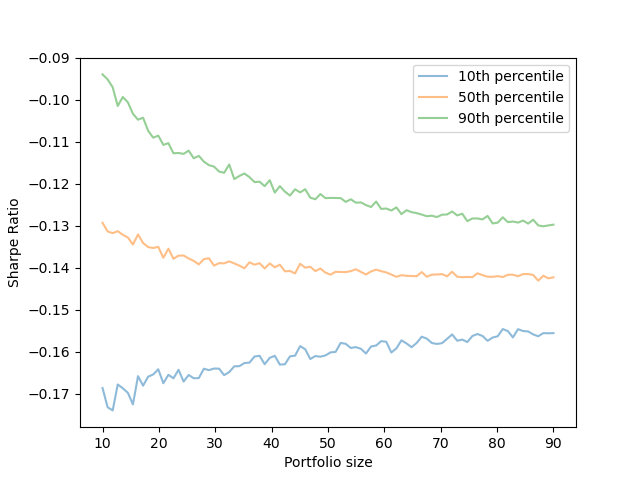}
        \caption{} 
        \label{fig:Portfolio_bear_3}
    \end{subfigure}
    \begin{subfigure}[b]{0.32\textwidth}
        \includegraphics[width=\textwidth]{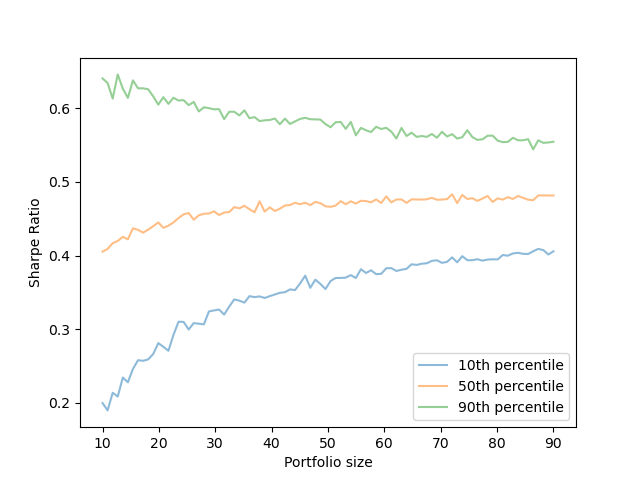}
        \caption{} 
        \label{fig:Portfolio_bull_1}
    \end{subfigure}
    \begin{subfigure}[b]{0.32\textwidth}
        \includegraphics[width=\textwidth]{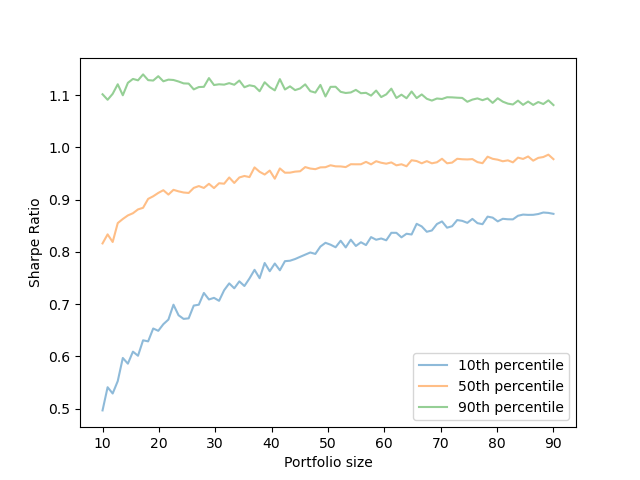}
        \caption{} 
        \label{fig:Portfolio_bull_2}
    \end{subfigure}
        \begin{subfigure}[b]{0.32\textwidth}
        \includegraphics[width=\textwidth]{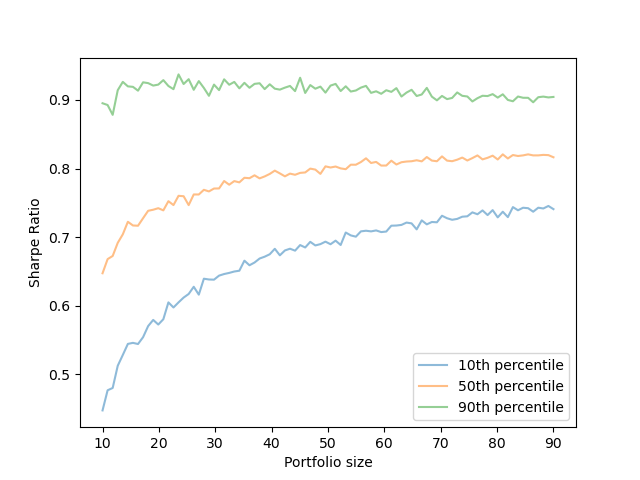}
        \caption{} 
        \label{fig:Portfolio_bull_3}
    \end{subfigure}
    \caption{Quantile values $Q_{\mathcal{S},k}(q)$ of the distribution of Sharpe ratios for $q \in \{ 0.1, 0.5, 0.9\}$, recorded as a function of the portfolio size $k$. $q=0.1, 0.5, 0.9$ represent low-skilled, median and high-skilled investors respectively. We display these functions for six one-year periods, comprising three bear markets (a) mid 2000 - mid 2001, (b)  2001 -  2002, (c)  2007  -  2008 and three bull markets (d) 2011 - 2012, (e) 2012 - 2013 (f) 2015 - 2016.}
    \label{fig:Portfolio_k_curves}
\end{figure*}

Figure \ref{fig:Portfolio_k_curves} shows the recorded Sharpe ratio quantiles at varying portfolio sizes during six yearly periods in the market; three sampled from equity bear markets (\ref{fig:Portfolio_bear_1}, \ref{fig:Portfolio_bear_2} and \ref{fig:Portfolio_bear_3}) and three sampled from bull markets (\ref{fig:Portfolio_bull_1}, \ref{fig:Portfolio_bull_2} and \ref{fig:Portfolio_bull_3}). There are several key findings in the figures. First, we see substantial consistency in behaviour across the different periods. As portfolios get larger, there is increased marginal cost and benefit for skilful investors and less skilled investors, respectively. Second, we see significantly lower Sharpe ratios during the bear market periods, due to reduced returns and increased volatility. We remark that in this paper, portfolios are constructed in a long-only capacity, and skilled long-short investors who can successfully identify strong shorting opportunities during bear markets may in fact outperform during equity crises. 
Third, we observe the greatest shifts in recorded quantile values for small portfolio size $k$, with this effect reducing as more securities are sequentially added to equity portfolios. For small cardinality portfolios, we see a massive range in expected Sharpe ratios based on investor skill levels; this range is reasonably consistent across both bull and bear market dynamics. Finally, we see an interesting distinction between the bull and bear markets: in the bull periods, there is a greater marginal benefit of increasing portfolio size for low skilled than skilled investors - as seen by the greater change over time in the bottom vs top curve. However, in bear markets, the opposite behaviour is observed.

\begin{figure*}
    \centering
        \begin{subfigure}[b]{0.32\textwidth}
        \includegraphics[width=\textwidth]{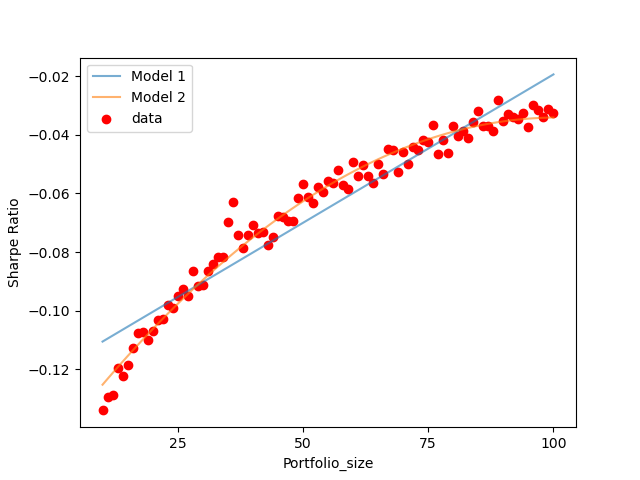}
        \caption{}
        \label{fig:Portfolio_bear_1_reg}
    \end{subfigure}
        \begin{subfigure}[b]{0.32\textwidth}
        \includegraphics[width=\textwidth]{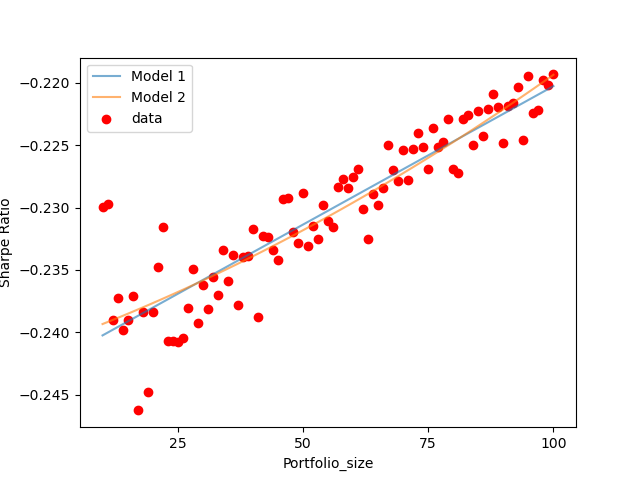}
        \caption{}
        \label{fig:Portfolio_bear_2_reg}
    \end{subfigure}
            \begin{subfigure}[b]{0.32\textwidth}
        \includegraphics[width=\textwidth]{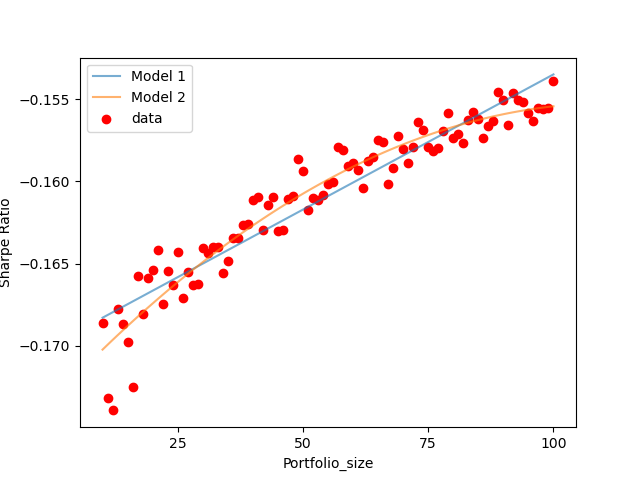}
        \caption{}
        \label{fig:Portfolio_bear_3_reg}
    \end{subfigure}
    \begin{subfigure}[b]{0.32\textwidth}
        \includegraphics[width=\textwidth]{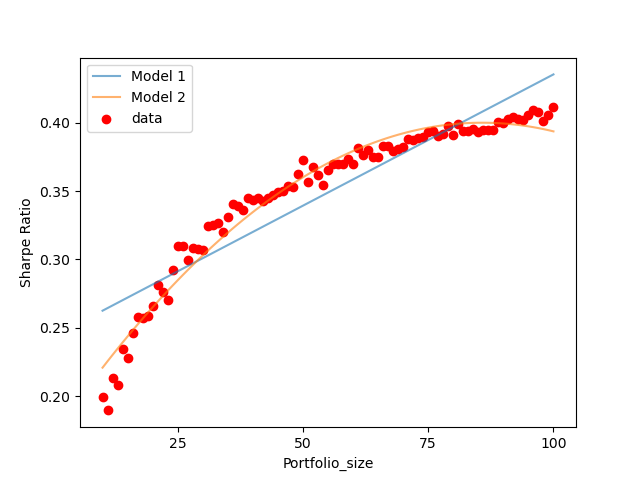}
        \caption{}
        \label{fig:Portfolio_bull_1_reg}
    \end{subfigure}
        \begin{subfigure}[b]{0.32\textwidth}
        \includegraphics[width=\textwidth]{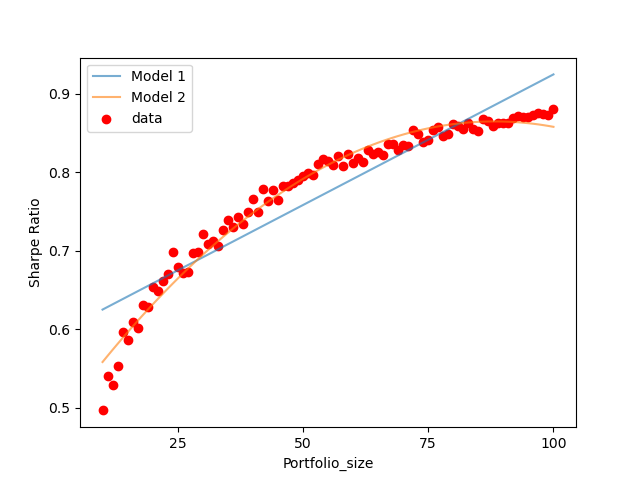}
        \caption{}
        \label{fig:Portfolio_bull_2_reg}
    \end{subfigure}
            \begin{subfigure}[b]{0.32\textwidth}
        \includegraphics[width=\textwidth]{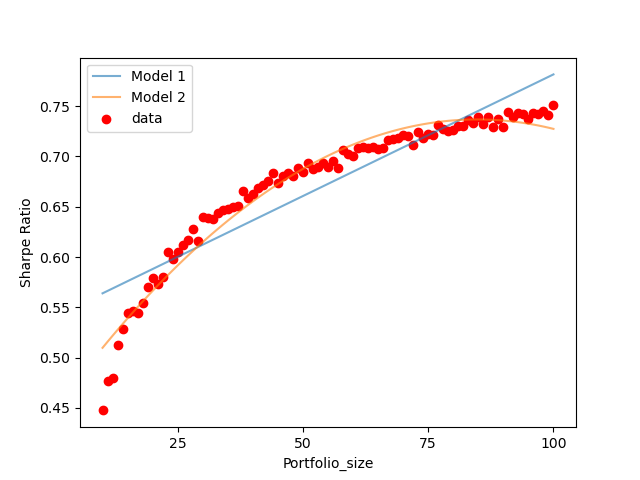}
        \caption{}
        \label{fig:Portfolio_bull_3_reg}
    \end{subfigure}
    \caption{Linear and quadratic regression fits for the Sharpe quantile $Q_{\mathcal{S},k}(q)$ against $k$, displayed for low-skilled investors $q=0.1$. We display these regressions for six one-year periods, comprising three bear markets (a) mid 2000 - mid 2001, (b)  2001 -  2002, (c)  2007  -  2008 and three bull markets (d) 2011 - 2012, (e) 2012 - 2013 (f) 2015 - 2016. We observe the bull markets (second row) observe a more pronounced non-linear curvature.}
    \label{fig:Portfolio_regressions}
\end{figure*}

Next, we examine the regression fits between portfolio size and Sharpe ratio, with a focus on less skilled investors ($q=0.1$) over different periods, choosing the same three bear market and three bull periods as Figure \ref{fig:Portfolio_k_curves}. Figure \ref{fig:Portfolio_regressions} presents an interesting finding: the bull market periods (\ref{fig:Portfolio_bull_1_reg}, \ref{fig:Portfolio_bull_2_reg} and \ref{fig:Portfolio_bull_3_reg}) display substantially more visible curvature (a stronger quadratic fit) than the bear markets (\ref{fig:Portfolio_bear_1_reg}, \ref{fig:Portfolio_bear_2_reg} and \ref{fig:Portfolio_bear_3_reg}). In fact, Figure \ref{fig:Portfolio_bear_2_reg} is one of the most strongly linear fits in the entire analysis across all years and skill levels, discussed further in the supplementary material. In the bear periods, this suggests there is a reasonably steady marginal benefit in portfolio diversification (with respect to size) for unskilled investors, whereas in bull markets, the benefit is strongest for small increases in $k$, with diminishing returns observed thereafter. In fact, all three bull markets exhibit clear diminishing returns beyond a portfolio of $\sim 50$ stocks. That is, for a less skilled investor, there is reduced relative benefit in holding a portfolio larger than 50 stocks during a bull market.

The supplementary material contains substantially more details regarding our regression experiments - here we present a brief overview. First, for every yearly period and quantile level, the linear regression coefficient $\hat{\beta}^1_1$ is highly significant (even correcting for multiple hypotheses). For the bottom decile investors ($q=0.1$), $\hat{\beta}^1_1$ is positive in every year, indicating that less competent investors should overwhelmingly hold more stocks in their portfolio. For the top decile investors, $\hat{\beta}^1_1$ is always negative, implying that for the most competent decile of investors, holding a larger number of equities will lead to lower expected risk-adjusted returns. For median investors, an interesting story emerges: $\hat{\beta}^1_1$ is always highly significant, but positive in some years and negative in others. Moreover, there is a strong association between the $\hat{\beta}^1_1$ value being negative, and poor performance of the S\&P 500 index. The years where $\hat{\beta}^1_1<0$ correspond to 2002, 2007, 2008, 2011, 2015 and 2018. The respective returns of the S\&P 500 index during those years was -23.37\%, 3.5\%, -38.5\%, -0.01\%, -0.73\% and -6.24\%. During the other years, where $\hat{\beta}^1_1>0$, the S\&P 500 was strongly positive - averaging double digit positive returns. This presents a dynamic strategy for median investors seeking to improve their expected risk-adjusted returns: hold fewer, high-conviction investments during times where the index is underperforming. These periods in the market are most likely when investors may be most unsure of their portfolio positions; the natural tendency would be to hold a larger number of securities to naively diversify as much as possible.

\section{Continuous analysis}
\label{sec:continuousanalysis}

\begin{figure*}
    \centering
\includegraphics[width=0.99\textwidth]{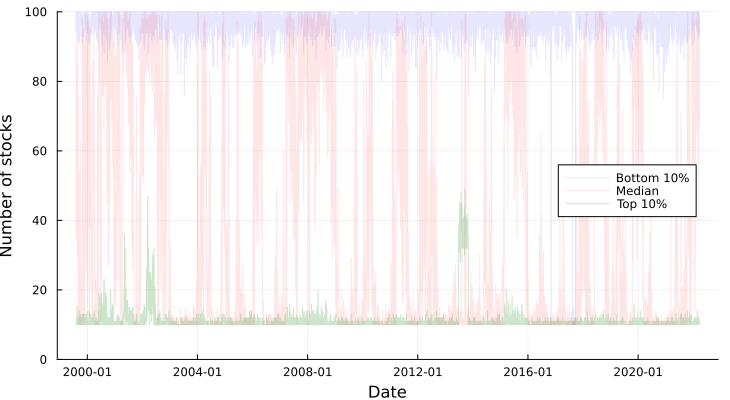}
    \caption{Continuously varying optimal portfolio size $k_0$ for the Sharpe quantile function $Q_{\mathcal{S},k}(q)$ for $q=0.1, 0.5, 0.9$ quantiles, representing low-skilled, median and high-skilled investors respectively.}
    \label{fig:threecurves}
\end{figure*}

\begin{figure}
    \centering
\includegraphics[width=0.49\textwidth]{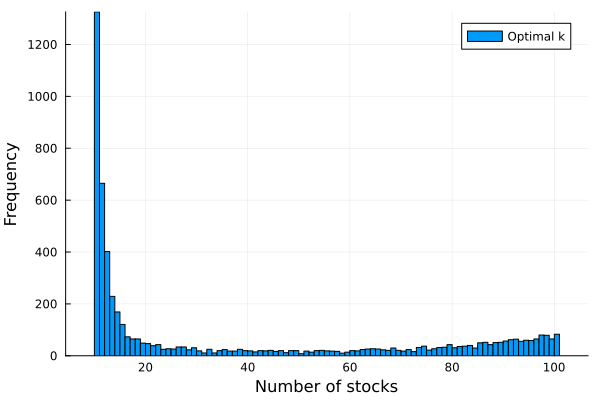}
    \caption{Histogram of the time-varying optimal portfolio size $k_0$ for the median Sharpe ratio ($q=0.5$).}
    \label{fig:k_median_hist}
\end{figure}

In this section, we aim to investigate trends in the representative Sharpe ratios on a continuous basis. Thus, we investigate a shorter interval of $P=90$ days, and perform analogous experiments as in Section \ref{sec:annualanalysis} with Sharpe ratios sampled over rolling intervals $[1,P],[2,P+1],...,[T-P+1,T]$ with $T=5788$. Thus, our functions $k \mapsto Q_{\mathcal{S},k}(q)$ may be computed on a continuous time-varying basis. For all three $q$, we compute the raw maximum $k_0$, which now becomes a time-varying function of $t=P,...,T$. All three functions are displayed in Figure \ref{fig:threecurves}. For $q=0.1$ and $0.9$ respectively, the optimal $k_0$ is overwhelmingly close to the largest and smallest possible portfolio cardinalities of $100$ and $10$ respectively. For the median investor ($q=0.5$) the behaviour of $k_0$ is less clear. To further elucidate the median situation, we plot the histogram of $k_0$ for the median ($q=0.5$) in Figure \ref{fig:k_median_hist}.

\subsection{Continuous analysis results}

Figure \ref{fig:threecurves} shows the optimal number of stocks in a portfolio for investors in the top 10\%, median and bottom 10\% of risk-adjusted returns (measured by the Sharpe ratio) over time. Our findings demonstrate that low-skilled investors overwhelmingly should select nearly 100 stocks (our experiment's upper bound); highly skilled investors should mostly select close to 10 stocks (our lower bound), while the median investor has a more complicated time-varying distribution of ideal portfolio size. The results for the high and low skilled investors are consistent with general investment sensibility, which is that investors with higher conviction in their positions can perform better when holding fewer stocks. Investors who fail to successfully identify strong performing equities do best with larger equity portfolios, as the larger sample  leads to superior diversification benefits and some mean reversion. It is difficult to extract a clear takeaway from the median investors' optimal portfolio size, other than the high variance in the optimal number makes it difficult to correctly size a portfolio in a non-dynamic (not time-varying) manner. We further investigate this phenomenon in Figure \ref{fig:k_median_hist}, where we examine the distribution over time of the optimal number of stocks in equity portfolios for median investors. The distribution plot demonstrates that a low number of equities is typically favoured (rather surprisingly), but there is no clear temporal structure as to when this precisely happens. However, in examining the behaviour of highly skill investors, it is worth noting that their optimal $k$ departs from close to the minimal number stocks at particular periods, such as around the start of 2014. This period may correspond to a market where many equities have reasonably strong returns, and holding a larger portfolio may reduce the portfolio's volatility.

\section{Conclusion}

This work tackles a familiar problem, with a unique twist. Portfolio construction is not usually addressed with the concept of investor ability in mind. In this work, we quantitatively investigate optimal portfolio construction as investor abilities vary. There are three key takeaways from our work. First, holding a larger number of equities in an investment portfolio leads to superior performance for poor investors, while holding fewer equities drives superior performance for strong investors. Second, we show that for weak investors, portfolio Sharpe ratios (as a function of portfolio cardinality) exhibit more quadratic curvature in bull markets than in bear markets, the latter being close to linear in some periods. Finally, our time-varying analysis demonstrates that median investors have a less obvious distribution with varying optimal portfolio size over time.

The findings in our work may have meaningful implications for a range of investor and fund allocators (those allocating capital to investment management companies) in various scenarios. Principally, one main takeaway of this work is that a high-skilled investor should make a small collection of high conviction investments rather than holding the entire index, which would amount to a largest possible choice of portfolio size $k$. By contrast, our findings encourage a low-skilled investor to hold a maximal $k$, which in practice may resort to buying the entire index (or ETF thereof) - providing cheap access to market beta.

Many investment managers such as those running ``private equity style'' public equity portfolios, insist on constructing portfolios with fewer stocks and higher conviction bets. Our work confirms this strategy for highly skilled investors, and furthermore, shows that this strategy is robust through varying market conditions. Our analysis suggests that investors with deteriorating performance should consider holding a larger number of equities in their portfolios, as a means of diversification benefit and accepting some mean-reversion in returns. One way this could be managed by asset consultants or funds management allocators would be insisting that fund investment policy statements are more flexible on portfolio cardinality, and pressuring fund managers to increase the number of securities held after (or during) periods of poor investment performance. A similar strategy could be taken more proactively by investors, holding a larger number of securities when they experience market conditions where their conviction is lower. This could be driven lower by less business-specific knowledge, or due to difficult economic conditions. Finally, our results show that optimal portfolio cardinality varies meaningfully for median investors over time. This optimal number of securities could be associated with specific macroeconomic conditions (or market regimes), and the best portfolio size could be estimated in a time-varying manner.

\section{Data availability statement}

All data are sourced from Bloomberg.

\clearpage
\onecolumn

\section{Supplementary material}

\section{Regression experiments}

In this section, we provide further details regarding the regression experiments described in the manuscript. As described in the main body of the manuscript, we first obtain our quantile Sharpe ratio values $Q_{\mathcal{S},k}(q)$, for three quantiles $q=0.1,0.5,0.9$, representative of a typical risk-adjusted return obtained by a typical low-skilled, medium and high-skilled investor respectively. These are computed for each one-year period of our full window of analysis. We are interested in these quantile values as a function of $k$, expressed mathematically as 
\begin{align}
\label{eq:nonpenalisedmax_supp}
   \{10,11,...,100\} \to \mathbb{R}; k \mapsto Q_{\mathcal{S},k}(q).
\end{align}
There is such a curve for each quantile $q=0.1,0.5,0.9$ and for every one-year period. We fit two regression models to each of the three curves, one linear, and one quadratic, notated as follows:
\begin{align}
\label{eq:linearmodel}
Q_{\mathcal{S},k}(q) &\sim \beta^1_0 + \beta^1_1 k,  \\
Q_{\mathcal{S},k}(q) &\sim \beta^2_0 + \beta^2_1 k + \beta^2_2 k^2.
\label{eq:quadraticmodel_supp}
\end{align}
Across three values of $q$ and 22 year-length periods of analysis, we have 66 different curves $k \mapsto Q_{\mathcal{S},k}(q)$, for each of which two models are fit. We investigate which of these two models provides a better fit across multiple criteria.

We say that the quadratic model (\ref{eq:quadraticmodel_supp}) provides an unambiguously better fit than the linear model (\ref{eq:linearmodel}) if the following four conditions are met:
\begin{itemize}
    \item the quadratic model has a lower Akaike information criterion (AIC);
    \item the quadratic model has a lower Bayesian information criterion (BIC);
    \item the quadratic model has a higher adjusted $R^2$ (Adj $R^2$);
    \item the quadratic model's fitted $\hat{\beta}^2_2$ coefficient is deemed highly significant with $p<\frac{0.05}{66}$.
\end{itemize}
As we are simultaneously testing 66 hypotheses for the last condition, we use a conservative Bonferroni approach for the last condition. Given $m$ hypothesis tests $H_{0,i}$ vs $H_{1,i}$ with $p$-values $P_i$ for $i=1,...,m$ and a significance threshold of $\alpha$, this approach rejects any $H_{0,i}$ with $P_i<\frac{\alpha}{m}$ to ensure that the probability of at least one false rejection is at most $\alpha$. For robustness, we also used a threshold of $\alpha=0.01$ and obtained identical results to $\alpha=0.05$.

We display selected results and statistical parameters in Tables \ref{tab:Lower},\ref{tab:Middle} and \ref{tab:Upper}, for yearly experiments on low-skilled, medium and high-skilled investors respectively ($q=0.1, 0.5, 0.9$, respectively). Each table contains a summary of the linear model (\ref{eq:linearmodel}) on the left and the quadratic model (\ref{eq:quadraticmodel_supp}) on the right. We include the AIC, BIC, adjusted $R^2$ and the final coefficient of each model (scaled for readability). We first remark that we have deliberately omitted any $p$-values for conciseness and readability. The $p$-values for the linear coefficient $\hat{\beta}^1_1$ of the linear model (\ref{eq:linearmodel}) are highly significant in every year and each value of $q$, so these are safely omitted. The $p$-values for the quadratic coefficient $\hat{\beta}^2_2$ of the quadratic model (\ref{eq:quadraticmodel_supp}) are highly significant in all but eight total rows across the three tables - we note these exceptions and record their $p$-values in the table captions.

For almost every year (row of data) across the three tables, the quadratic model provides an unambiguously better fit. With the exception of just the eight aforementioned rows, the quadratic model satisfies all of the above conditions (lower AIC, lower BIC, higher adj $R^2$, and highly significant leading term) to declare it a better fit. These eight anomalous rows are marked in red throughout the three tables. Examining these rows more closely, we have some notable findings. First, the AIC for the quadratic model is always lower than the linear model throughout our experiments, without exception. Second, the adjusted $R^2$ for the quadratic model is always higher, without exception. Despite this, however, we note that there are six instances, entirely contained within the the aforementioned eight rows, where the adjusted $R^2$ of the two models are \emph{close}, within 0.02 to be precise. In all other cases, the adjusted $R^2$ scores differ by at least 0.05. These close differences in adjusted $R^2$ are marked in blue in the tables. Third, there are four instances, again entirely contained within the aforementioned six, in which the BIC for the linear model is lower than that of the quadratic. We mark these in purple.

\begin{figure*}
    \centering
    \begin{subfigure}[b]{0.32\textwidth}
        \includegraphics[width=\textwidth]{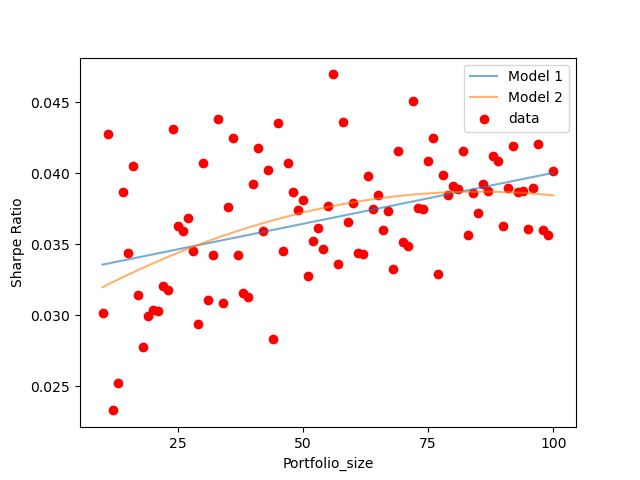}
        \caption{} 
        \label{fig:50_504}
    \end{subfigure}
    \begin{subfigure}[b]{0.32\textwidth}
        \includegraphics[width=\textwidth]{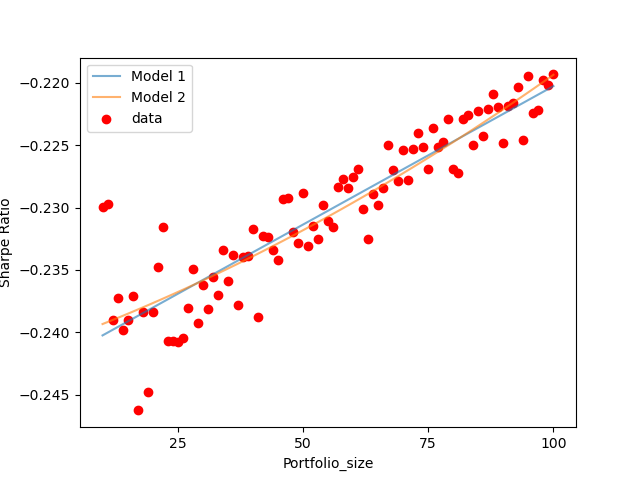}
        \caption{} 
        \label{fig:10_756}
    \end{subfigure}
        \begin{subfigure}[b]{0.32\textwidth}
        \includegraphics[width=\textwidth]{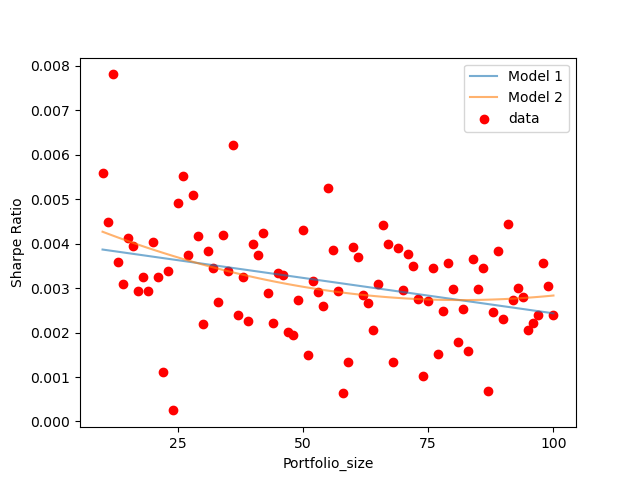}
        \caption{} 
        \label{fig:50_3024}
    \end{subfigure}
    \begin{subfigure}[b]{0.32\textwidth}
        \includegraphics[width=\textwidth]{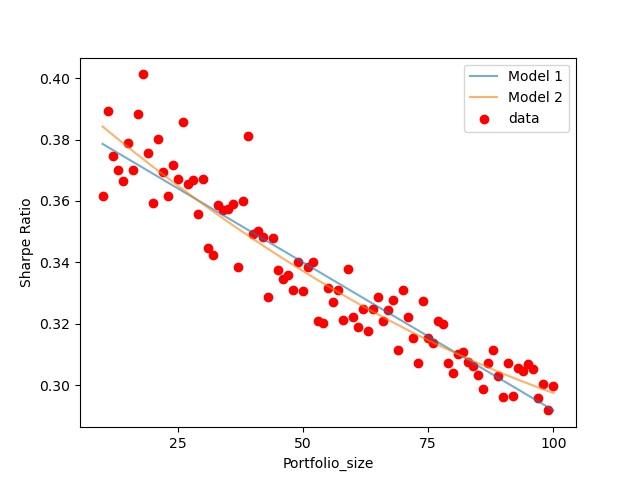}
        \caption{} 
        \label{fig:90_252}
    \end{subfigure}
    \begin{subfigure}[b]{0.32\textwidth}
        \includegraphics[width=\textwidth]{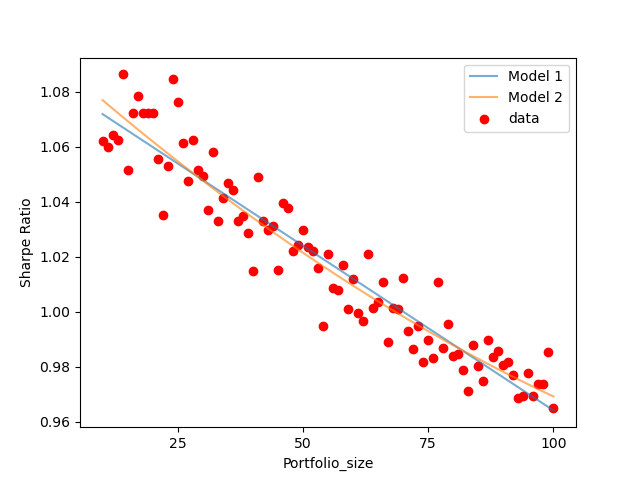}
        \caption{} 
        \label{fig:90_5040}
    \end{subfigure}
        \begin{subfigure}[b]{0.32\textwidth}
        \includegraphics[width=\textwidth]{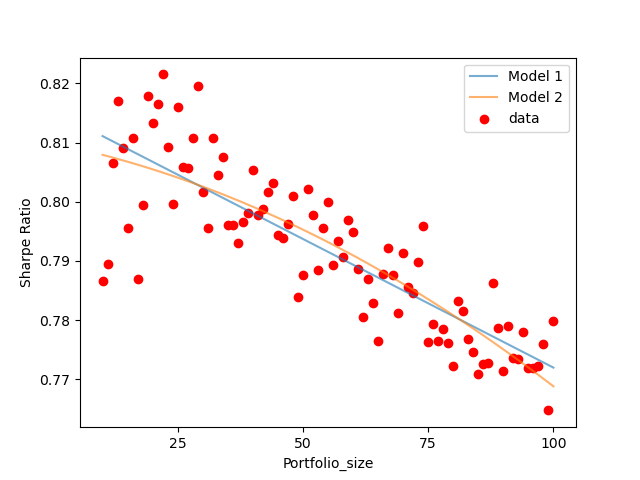}
        \caption{} 
        \label{fig:90_5544}
    \end{subfigure}
    \caption{Linear and quadratic regression fits for the Sharpe quantile $Q_{\mathcal{S},k}(q)$ against $k$. We select the subset of six anomalous rows detailed in the text, corresponding to (a) $q=0.5$ for 2000-2001, (b) $q=0.1$ for 2001-2002, (c) $q=0.5$ for 2007-2008, (d) $q=0.9$ for 2011-2012, (e) $q=0.9$ for 2012-2013, (f) $q=0.9$ for 2015-2016. Figure 1(b) is identical to Figure 2(b) of the main manuscript. These six anomalous examples are the most linear fits among all the recorded experiments.}
    \label{fig:Linearfits}
\end{figure*}

\begin{figure*}
    \centering
    \begin{subfigure}[b]{0.32\textwidth}
        \includegraphics[width=\textwidth]{Portfolio_size_regression_d10_3528.png}
        \caption{} 
        \label{fig:10_3528}
    \end{subfigure}
    \begin{subfigure}[b]{0.32\textwidth}
        \includegraphics[width=\textwidth]{Portfolio_size_regression_d10_4284.png}
        \caption{} 
        \label{fig:10_4284}
    \end{subfigure}
    \begin{subfigure}[b]{0.32\textwidth}
        \includegraphics[width=\textwidth]{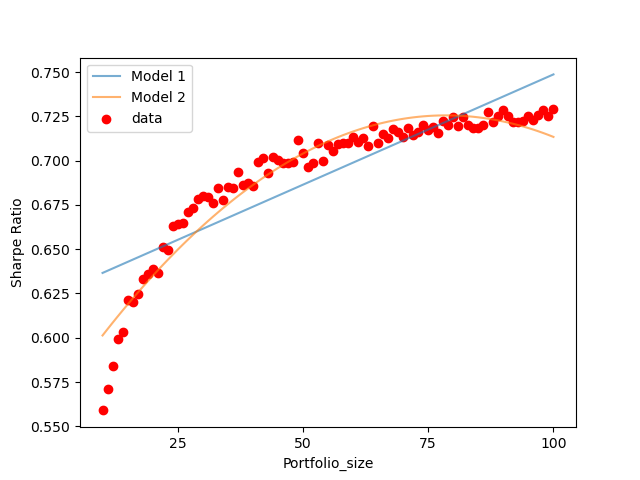}
        \caption{} 
        \label{fig:50_1008}
    \end{subfigure}
    \begin{subfigure}[b]{0.32\textwidth}
        \includegraphics[width=\textwidth]{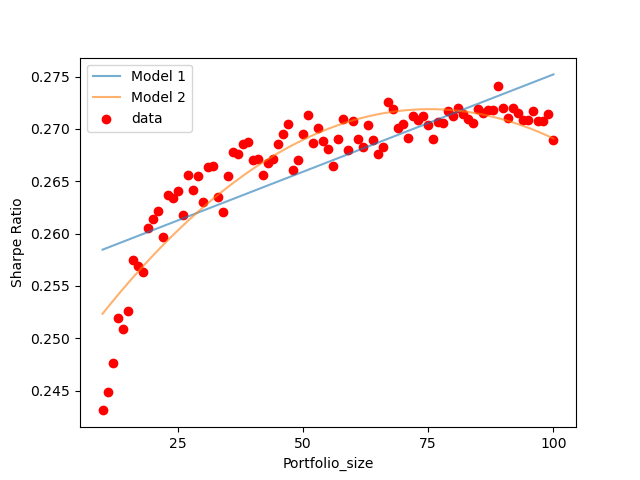}
        \caption{} 
        \label{fig:50_2772}
    \end{subfigure}
        \begin{subfigure}[b]{0.32\textwidth}
        \includegraphics[width=\textwidth]{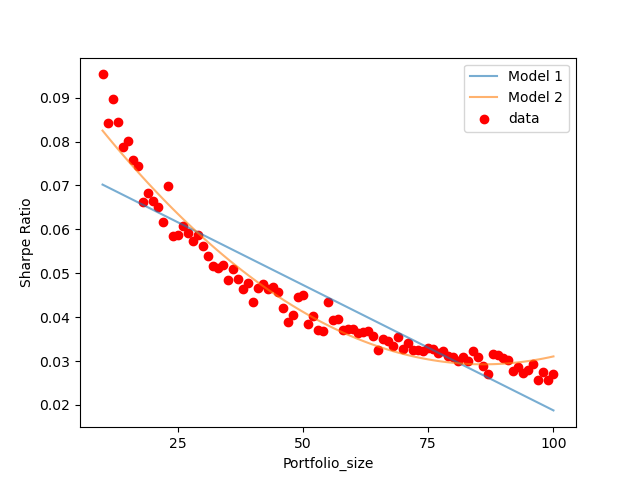}
        \caption{} 
        \label{fig:90_3024}
    \end{subfigure}
        \begin{subfigure}[b]{0.32\textwidth}
        \includegraphics[width=\textwidth]{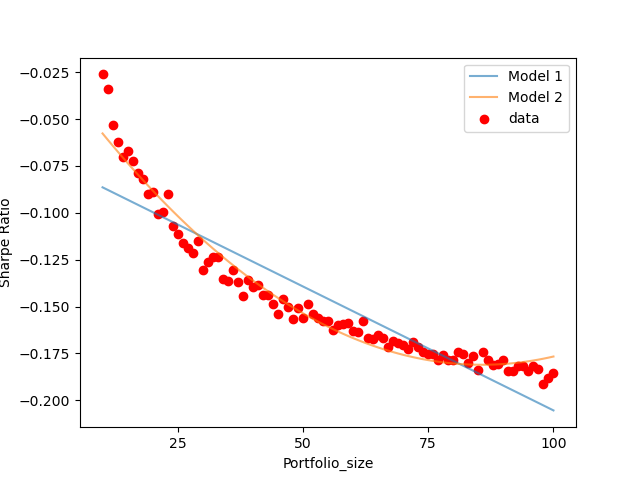}
        \caption{} 
        \label{fig:90_4788}
    \end{subfigure}
    \caption{Linear and quadratic regression fits for the Sharpe quantile $Q_{\mathcal{S},k}(q)$ against $k$. We select rows where the quadratic fit is particularly apparent, corresponding to (a) $q=0.1$ for 2012-2013, (b) $q=0.1$ for 2015-2016, (c) $q=0.5$ for 2002-2003, (d) $q=0.5$ for 2009-2010, (e) $q=0.9$ for 2010-2011, (f) $q=0.9$ for 2017-2018.}
    \label{fig:quadraticfits}
\end{figure*}

That is, we identify a set of four exceptions in BIC, where the linear model has a better score. These four are a subset of six exceptions in adjusted $R^2$, where the quadratic model has only a slightly better score. These six are a subset of eight total exceptional rows, where the quadratic model's final term is not highly significant, meaning we cannot reject the null hypothesis of no curvature. We plot the regressions for the subset of six (the blue-marked rows) in Figure \ref{fig:Linearfits}. In particular, Figure \ref{fig:10_756} is identical to Figure 2b of the main manuscript, and is an example where the $p$-value of $\hat{\beta}_2^2$ is $p=0.68$, failing to provide evidence to reject a lack of curvature. We can see this in the figure - it is a rather exceptional example where no curvature in the representative Sharpe value quantile curve is visually observed. By way of contrast, in Figure \ref{fig:quadraticfits} we show regression fits across $q=0.1,0.5,0.9$ where the quadratic fit is particularly apparent. In the manuscript, we comment and note that a more apparent quadratic fit (with greater curvature) for $q=0.1$ is more frequently seen for bull markets than bear markets

In the main body of the manuscript, we provide an ample discussion of the sign of the fitted linear coefficient $\hat{\beta}^1_1$. For completeness, we briefly repeat the primary results. For the bottom decile experiment ($q=0.1$), $\hat{\beta}^1_1$ is positive and highly significant in every year. For the top decile experiment ($q=0.9$), $\hat{\beta}^1_1$ is negative and highly significant in every year. For the median investor ($q=0.5$), $\hat{\beta}^1_1$ is always highly significant, but it is sometimes negative, and sometimes positive. There are six years where $\hat{\beta}^1_1<0$ for the $q=0.5$ quantile curve, corresponding to 2002, 2007, 2008, 2011, 2015 and 2018.


\section{Robustness of penalised optimum}

In the main body of the manuscript, we discuss two methods for selecting the appropriate portfolio size $k$, namely the \emph{raw optimum} $k_0$ for each quantile $q=0.1,0.5,0.9$ and year under consideration, and the penalised optimum $\hat{k}$. The latter only makes sense when $\hat{\beta}^1_1>0$, which is the case for every year with $q=0.1$ and certain years with $q=0.5$. We include the record of $k_0$ and, where applicable, $\hat{k}$, in Tables \ref{tab:Lower}, \ref{tab:Middle} and \ref{tab:Upper}. In addition, we include a brief scatter plot to demonstrate the robustness of the penalised optimum as an alternative strategy for investors compared to the raw optimum.

\begin{figure}
    \centering
\includegraphics[width=\textwidth]{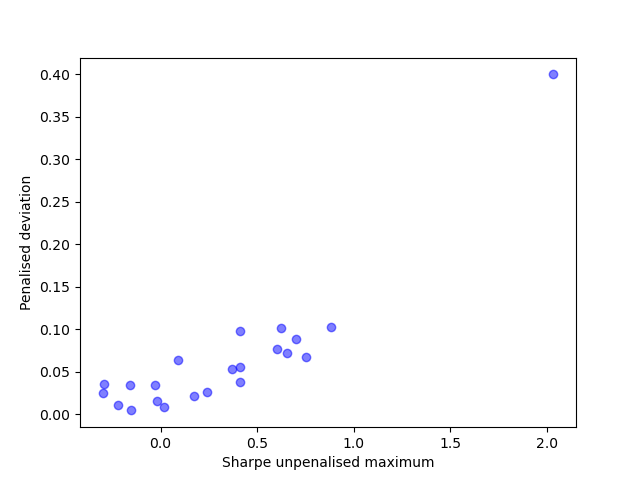}
    \caption{Scatter plot between unpenalised optimum Sharpe ratio and deviation with the penalised maximum, across all years for $q=0.1$. This serves as a check of robustness for our penalised optimum, that it yields a closely similar Sharpe ratio in all but one year.}
    \label{fig:Penalised_sharpe_deviation}
\end{figure}

In Figure \ref{fig:Penalised_sharpe_deviation}, we display the deviation in representative Sharpe ratios $\Delta=Q_{\mathcal{S},k_0}(q) - Q_{\mathcal{S},\hat{k}}(q)$ against the raw optimal Sharpe ratio $Q_{\mathcal{S},k_0}(q)$, all specifically for $q=0.1$, over all our yearly intervals. This deviation $\Delta$ is by definition always positive, and we can see that for most years, it is relatively small. This indicates that in performance terms, little is lost by using the penalised optimum for the less skilled investor. Specifically, all but one of the deviations $\Delta$ are at most $\sim 0.1$, with one exceptional value of $\sim 0.4$. That is, only a single outlier year is observed in which there is a substantial deviation between using the raw and penalised optimum portfolio size. Thus, our penalised optimisation framework may significantly reduce the number of stocks to be held with little difference in portfolio Sharpe ratio. This method could be of particular interest to investors who experience high transaction costs, or investors who are especially confident in their ability to identify strong performing stocks in an out-of-sample testing period.

\newpage

\begin{table}
\begin{center}
\begin{tabular}{c|cccc|cccc|cc}
\toprule
Year & AIC & BIC & Adj $R^2$ & $\hat{\beta}^1_1$ & AIC & BIC & Adj $R^2$ & $\hat{\beta}^2_2$ & $k_0$ & $\hat{k}$\\
\midrule
0 & -524.81 & -519.79 & 0.92 & 1.68 & -656.06 & -648.53 & 0.98 & -18.81 & 97 & 39 \\
1 & -617.89 & -612.87 & 0.92 & 1.01 & -735.93 & -728.40 & 0.98 & -11.01 & 89 & 36\\
\textcolor{red}{2} & -816.14 & \textcolor{purple}{-811.12} & \textcolor{blue}{0.82} & 0.22 & -816.43 & \textcolor{purple}{-808.89} & \textcolor{blue}{0.83} & \textcolor{red}{0.68$^1$} & 100 & 10 \\
3 & -383.96 & -378.94 & 0.82 & 2.34 & -519.36 & -511.83 & 0.96 & -41.06 & 99  & 41\\
4 & -400.97 & -395.95 & 0.86 & 2.51 & -554.99 & -547.46 & 0.97 & -38.37 & 100 & 52\\
5 & -335.67 & -330.64 & 0.85 & 3.48 & -472.46 & -464.92 & 0.97 & -53.66 & 97 & 44\\
6 & -391.93 & -386.90 & 0.85 & 2.55 & -532.20 & -524.67 & 0.97 & -39.60 & 97 & 35\\
7 & -529.73 & -524.71 & 0.87 & 1.30 & -646.14 & -638.60 & 0.97 & -17.82 & 96 & 44\\
8 & -916.79 & -911.77 & 0.89 & 0.16 & -952.76 & -945.23 & 0.92 & -1.45 & 100  & 49\\
9 & -651.59 & -646.57 & 0.84 & 0.58 & -764.65 & -757.12 & 0.95 & -9.06 & 94 & 35\\
10 & -567.97 & -562.95 & 0.83 & 0.87 & -697.57 & -690.04 & 0.96 & -14.80 & 98 & 37\\
11 & -707.67 & -702.65 & 0.85 & 0.44 & -822.23 & -814.70 & 0.96 & -6.68 & 97 & 36\\
12 & -434.72 & -429.70 & 0.84 & 1.92 & -572.37 & -564.84 & 0.97 & -31.18 &100 & 50\\
13 & -352.94 & -347.92 & 0.87 & 3.33 & -508.82 & -501.29 & 0.98 & -50.07 &100 & 42\\
14 & -388.93 & -383.91 & 0.84 & 2.46 & -530.99 & -523.46 & 0.97 & -40.36 & 95 & 47\\
15 & -616.62 & -611.59 & 0.90 & 0.94 & -678.37 & -670.83 & 0.95 & -9.20  & 100 & 50\\
16 & -385.34 & -380.32 & 0.83 & 2.42 & -517.74 & -510.20 & 0.96 & -40.56 & 100 & 44\\
17 & -119.32 & -114.30 & 0.85 & 11.37 & -283.27 & -275.74 & 0.98 & -182.37 & 98 & 36\\
18 & -655.75 & -650.73 & 0.90 & 0.76 & -720.83 & -713.29 & 0.95 & -7.55 &100 & 29\\
19 & -333.34 & -328.32 & 0.85 & 3.41 & -472.78 & -465.25 & 0.97 & -54.57 &100 & 46\\
20 & -807.47 & -802.45 & 0.84 & 0.25 & -912.47 & -904.93 & 0.95 & -3.78 & 99 & 36\\
21 & -404.42 & -399.40 & 0.86 & 2.40 & -560.46 & -552.93 & 0.97 & -37.74 & 98  & 42\\
\bottomrule
\end{tabular}
\caption{Regression results for models (\ref{eq:linearmodel}) on the left and (\ref{eq:quadraticmodel_supp}) on the right for the lower decile $q=0.1$. In the table above, $\hat{\beta}^1_1$ has been multiplied by $10^3$ while $\hat{\beta}^2_2$ has been multipled by $10^6$. Purple denotes instances where the linear model presents a superior BIC, while blue denotes instances where the adjusted $R^2$ values are close (though the quadratic model always presents a higher adjusted $R^2$). The anomalous $p$-value (marked in red) is $p=0.14$ for $^1$, which fails the significance threshold of the Bonferroni condition. We include the raw optimum $k_0$ and penalised optimum $\hat{k}$ for every year.}
\label{tab:Lower}
\end{center}
\end{table}

\begin{table}
\begin{center}
\begin{tabular}{c|cccc|cccc|cc}
\toprule
Year & AIC & BIC & Adj $R^2$ & $\hat{\beta}^1_1$ & AIC & BIC & Adj $R^2$ & $\hat{\beta}^2_2$ & $k_0$ & $\hat{k}$\\
\midrule
0 & -616.39 & -611.37 & 0.68 & 0.45 & -680.99 & -673.45 & 0.85 & -9.35 & 71 & 39  \\
\textcolor{red}{1} & -741.43 & \textcolor{purple}{-736.41} & \textcolor{blue}{0.17} & 0.07 & -742.51 & \textcolor{purple}{-734.98} & \textcolor{blue}{0.19} & \textcolor{red}{-1.19$^1$} & 56 & 56\\
2 & -692.10 & -687.08 & 0.74 & -0.34 & -789.21 & -781.68 & 0.91 & 6.97 & 11 \\
3 & -457.09 & -452.07 & 0.74 & 1.25 & -571.32 & -563.78 & 0.93 & -26.44 & 100 & 49\\
4 & -542.10 & -537.08 & 0.74 & 0.79 & -656.30 & -648.76 & 0.93 & -16.57 & 100 & 53\\
5 & -435.72 & -430.70 & 0.75 & 1.45 & -549.07 & -541.53 & 0.93 & -29.67 & 99 & 35\\
6 & -510.89 & -505.87 & 0.76 & 0.98 & -622.08 & -614.55 & 0.93 & -19.54 & 97 & 42\\
\textcolor{red}{7} & -732.35 & -727.32 & 0.25 & -0.09 & -738.25 & -730.72 & 0.30 & \textcolor{red}{1.98$^2$} & 14 \\
8 & -907.14 & -902.12 & 0.72 & -0.10 & -1012.1 & -1004.6 & 0.91 & 2.18 & 10\\
9 & -827.45 & -822.43 & 0.64 & 0.13 & -880.36 & -872.83 & 0.80 & -2.74 & 74 & 30\\
10 & -754.51 & -749.48 & 0.63 & 0.19 & -829.34 & -821.81 & 0.84 & -4.59 & 89 & 51\\
\textcolor{red}{11} & -972.07 & \textcolor{purple}{-967.05} & \textcolor{blue}{0.11} & -0.02 & -972.53 & \textcolor{purple}{-965.00} & \textcolor{blue}{0.12} & \textcolor{red}{0.30$^3$} &12 \\
12 & -593.15 & -588.13 & 0.74 & 0.58 & -696.26 & -688.73 & 0.92 & -12.20 &79 & 41\\
13 & -454.65 & -449.63 & 0.73 & 1.24 & -562.62 & -555.09 & 0.92 & -26.42 &98 & 40\\
14 & -553.19 & -548.16 & 0.68 & 0.64 & -619.37 & -611.84 & 0.85 & -13.33 &95 & 48\\
15 & -667.36 & -662.34 & 0.65 & -0.32 & -738.85 & -731.32 & 0.84 & 7.30 &11 \\
16 & -459.60 & -454.57 & 0.75 & 1.25 & -573.07 & -565.54 & 0.93 & -26.03 &100 & 44\\
17 & -164.39 & -159.37 & 0.80 & 7.32 & -311.84 & -304.31 & 0.96 & -139.61 &100 & 45\\
18 & -690.34 & -685.32 & 0.61 & -0.26 & -764.47 & -756.93 & 0.83 & 6.51 & 10 \\
19 & -436.59 & -431.57 & 0.71 & 1.30 & -540.46 & -532.93 & 0.91 & -28.89 & 99 & 56\\
20 & -1022.4 & -1017.4 & 0.14 & 0.01 & -1034.4 & -1026.8 & 0.25 & -0.53 & 21 & 21\\
21 & -479.88 & -474.86 & 0.75 & 1.12 & -604.92 & -597.39 & 0.94 & -23.82 & 89 & 42\\
\bottomrule
\end{tabular}
\caption{Regression results for models (\ref{eq:linearmodel}) on the left and (\ref{eq:quadraticmodel_supp}) on the right for the median $q=0.5$. In the table above, $\hat{\beta}^1_1$ has been multiplied by $10^3$ while $\hat{\beta}^2_2$ has been multipled by $10^6$. Purple denotes instances where the linear model presents a superior BIC, while blue denotes instances where the adjusted $R^2$ values are close (though the quadratic model always presents a higher adjusted $R^2$). The anomalous $p$-values (marked in red) are $p=0.085,0.005,0.12$ for $^{1,2,3}$, respectively, which fail the significance threshold of the Bonferroni condition. We include the raw optimum $k_0$ for every year and penalised optimum $\hat{k}$ for each year with $\hat{\beta}^1_1>0$.}
\label{tab:Middle}
\end{center}
\end{table}

\begin{table}
\begin{center}
\begin{tabular}{c|cccc|cccc|c}
\toprule
Year & AIC & BIC & Adj $R^2$ & $\hat{\beta}^1_1$ & AIC & BIC & Adj $R^2$ & $\hat{\beta}^2_2$ & $k_0$ \\
\midrule
\textcolor{red}{0} & -593.96 & -588.93 & \textcolor{blue}{0.89} & -0.96 & -599.93 & -592.40 & \textcolor{blue}{0.89} & \textcolor{red}{4.25$^1$} & 18 \\
1 & -580.19 & -575.16 & 0.91 & -1.19 & -675.84 & -668.31 & 0.97 & 12.84 & 11 \\
2 & -545.94 & -540.92 & 0.82 & -0.97 & -678.11 & -670.57 & 0.96 & 16.78 & 11\\
3 & -577.92 & -572.90 & 0.16 & -0.17 & -614.42 & -606.88 & 0.45 & -9.41 & 27\\
4 & -577.05 & -572.03 & 0.91 & -1.21 & -610.00 & -602.47 & 0.94 & 9.09 & 13\\
\textcolor{red}{5} & -563.57 & -558.55 & 0.53 & -0.44 & -570.05 & -562.51 & 0.57 & \textcolor{red}{-5.17$^2$} & 29\\
6 & -597.64 & -592.62 & 0.90 & -1.04 & -637.86 & -630.32 & 0.94 & 8.76 & 14\\
7 & -483.20 & -478.18 & 0.86 & -1.55 & -606.87 & -599.34 & 0.96 & 23.34 & 11 \\
8 & -761.90 & -756.87 & 0.84 & -0.31 & -903.22 & -895.69 & 0.97 & 5.19 & 10\\
9 & -677.27 & -672.25 & 0.84 & -0.51 & -785.81 & -778.28 & 0.95 & 7.79 & 13\\
10 & -691.96 & -686.94 & 0.88 & -0.54 & -783.67 & -776.14 & 0.96 & 6.87 & 14\\
11 & -652.89 & -647.87 & 0.84 & -0.57 & -778.10 & -770.56 & 0.96 & 9.21 & 10\\
12 & -620.10 & -615.08 & 0.89 & -0.86 & -678.02 & -670.49 & 0.94 & 8.83  & 13\\
13 & -604.01 & -598.99 & 0.68 & -0.48 & -617.98 & -610.45 & 0.73 & -5.57 & 18\\
14 & -507.30 & -502.28 & 0.88 & -1.52 & -614.98 & -607.45 & 0.96 & 19.77 & 12\\
15 & -468.43 & -463.41 & 0.85 & -1.66 & -626.37 & -618.84 & 0.97 & 26.61 & 10\\
16 & -603.98 & -598.96 & 0.27 & -0.20 & -624.71 & -617.17 & 0.43 & -6.53 & 25\\
17 & -224.42 & -219.40 & 0.57 & 3.04 & -332.62 & -325.08 & 0.87 & -93.66 & 89\\
18 & -499.29 & -494.27 & 0.84 & -1.32 & -627.24 & -619.70 & 0.96 & 21.52 & 10\\
\textcolor{red}{19} & -575.33 & -570.31 & \textcolor{blue}{0.91} & -1.20 & -578.47 & -570.94 & \textcolor{blue}{0.91} & \textcolor{red}{3.81$^3$} & 14\\
20 & -768.78 & -763.76 & 0.83 & -0.29 & -892.71 & -885.17 & 0.96 & 4.86  & 10 \\
\textcolor{red}{21} & -633.86 & \textcolor{purple}{-628.84} & \textcolor{blue}{0.71} & -0.43 & -635.62 & \textcolor{purple}{-628.09} & \textcolor{blue}{0.72} & \textcolor{red}{-2.37$^4$} & 22 \\ 
\bottomrule
\end{tabular}
\caption{Regression results for models (\ref{eq:linearmodel}) on the left and (\ref{eq:quadraticmodel_supp}) on the right for the upper decile $q=0.9$. In the table above, $\hat{\beta}^1_1$ has been multiplied by $10^3$ while $\hat{\beta}^2_2$ has been multipled by $10^6$. Purple denotes instances where the linear model presents a superior BIC, while blue denotes instances where the adjusted $R^2$ values are close (though the quadratic model always presents a higher adjusted $R^2$). The anomalous $p$-values (marked in red) are $p$-values $p=0.005,0.004,0.02,0.06$ for $^{1,2,3,4}$, respectively, which fail the significance threshold of the Bonferroni condition. We include the raw optimum $k_0$ for every year, while $\hat{k}$ is not applicable as $\hat{\beta}^1_1$ is always negative.
}
\label{tab:Upper}
\end{center}
\end{table}

\clearpage

\bibliography{_references}
\bibliographystyle{eplbib.bst}

\end{document}